\documentstyle[twoside,fleqn,espcrc2,epsfig]{article}

\newcommand{\AmS}{{\protect\the\textfont2
  A\kern-.1667em\lower.5ex\hbox{M}\kern-.125emS}}

\title{Critical correlators of three-dimensional gauge theories at finite 
temperature: exact results from universality }

\author{R. Fiore\address{Dipartimento di Fisica, Universit\`a della Calabria,\\
 Rende, I--87030 Cosenza, Italy},
F. Gliozzi and P. Provero\address{Dipartimento di Fisica
Teorica dell'Universit\`a di Torino\\
 via P.Giuria 1, I--10125 Torino, Italy}
        \thanks{presented by P. Provero}}
       
\begin{document}

\begin{abstract}
According to the Svetitsky-Yaffe conjecture, a three-dimensional gauge theory 
undergoing a
continuous deconfinement transition is in the same universality class as a 
two-dimensional
statistical model with order parameter taking values in the center of the 
gauge group. This allows
us to use conformal field theory techniques to evaluate exactly various 
correlation functions at the
critical point. In particular, we show that the plaquette operator of the 
gauge theory is mapped into
the energy operator of the dimensionally reduced model. The plaquette 
expectation value in
presence of static sources for three-dimensional $SU(2)$ and $SU(3)$ 
theories at 
the deconfinement
temperature can be exactly evaluated, providing some new insight about the 
structure of the color
flux tube in mesons and baryons. 
\end{abstract}

\maketitle

\noindent
Universality applied to the finite temperature 
deconfinement transition of gauge 
theories takes the form of the Svetitsky-Yaffe conjecture
\cite{sy}: a $(d+1)$-dimensional gauge theory with a second
order finite temperature deconfinement transition 
is in the same universality class as the
$d$-dimensional spin model with symmetry group of the order parameter given 
by the center of the gauge group (provided this spin model has a second order 
phase transition as well).
The aim of this work is to exploit universality arguments to 
investigate the physics of the deconfinement transition. 
The key point is that correlation functions at criticality are universal:
when the Svetitsky-Yaffe conjecture applies,
we can predict that the gauge theory critical correlators will coincide with
the ones of the spin model.\\
Now consider a $(2+1)$-dimensional gauge theory. The equivalent
spin model will be described, at the critical point, by a two-dimensional
conformal field theory, whose correlation functions are in principle
known exactly. Therefore universality gives us access to exact critical
correlators for $(2+1)$-dimensional gauge theories at the deconfinement
point.
\vskip0.3cm\noindent
Let us focus on an explicit example, $(2+1)$-dimensional $SU(2)$ gauge
theory, which the Svetitsky-Yaffe conjecture predicts to be in the same 
universality class as the two-dimensional Ising model.
The values of the critical indices for the gauge theory deconfinement 
transition have in fact been shown  
to coincide to high accuracy with the known ones
of the Ising model\cite{teper}.\\
To compute critical correlators of the gauge theory, we need a mapping 
between observables of the gauge theory and of the Ising model. The first
item in this mapping is the correspondence between the respective order
parameters: the Polyakov loop will correspond to the spin operator of the 
Ising model.\\
Consider now the
plaquette operator. 
Symmetry considerations suggest to assume
that the plaquette operator is mapped
into a linear combination of the identity and energy operators in the Ising 
model conformal algebra:
\begin{equation}
\Box \to c_1 1+c_\epsilon \epsilon\label{plaqope}
\end{equation}
In principle, one should add all the secondary fields belonging to the 
conformal families of the $1$ and $\epsilon$ operators, whose contributions
are expected to disappear at large distances. Eq. (\ref{plaqope}) contains
the two leading terms of the operator product expansion of the plaquette 
in terms of local fields of the Ising model. 
\vskip0.3cm\noindent
The simplest correlator to be studied 
to verify our assumption
is just the plaquette expectation value,
or rather its finite size scaling behavior. We consider a $(2+1)$-dimensional 
rectangular lattice of sides $L_1$, $L_2$, $L_t$ with $L_t\ll L_1,L_2$
and periodic boundary conditions on all directions. Then we tune the coupling 
$\beta$ to the critical value $\beta_c(L_t)$ (precise estimates are available
in the literature \cite{teper}), and we measure the plaquette expectation 
value $\Box(L_1,L_2)$ as a function of the lattice sides $L_1$ and $L_2$.\\
Universality gives us the following prediction for this quantity:
\begin{equation}
\Box(L_1,L_2)=c_{1}+c_{\epsilon}\frac{F(\tau)}{\sqrt{L_1 L_2}}
+O\left(\frac{1}{L_1 L_2}\right)\label{plaqfse}
\end{equation}
Here $c_1$ and $c_\epsilon$ are non universal constants whose value will 
depend on $L_t$ and on the kind of plaquette we consider (timelike or 
spacelike). The coefficient of $c_{\epsilon}$ in Eq.(\ref{plaqfse}) is 
the energy expectation value  for the 
two-dimensional Ising model 
on a torus at the critical point (see {\em e.g.} \cite{id});
$\tau=i\frac{L_1}{L_2}$ is the modular parameter of the torus, and
$F(\tau)$ is expressed in terms of Dedekind $\eta$ and Jacobi $\theta$ 
functions (see \cite{su2} for the explicit expression). 
The $O(1/L_1L_2)$ term comes from secondary fields: Eq.(\ref{plaqfse})
will be valid for asymptotically large lattices.\\
To verify the correctness of this prediction, 
we need to include all the data, coming from lattices
of different shapes, in a single fit. Let us therefore define an ``effective
area'' that incorporates the $\tau$ dependence predicted by the Ising model:
\begin{equation}
\alpha(L_1,L_2)=\frac{L_1 L_2}{F^2\left(i\frac{L_1}{L_2}\right)}
\end{equation}
Eq. (\ref{plaqfse}) predicts a linear dependence of $\Box(L_1,L_2)$
on $1/\sqrt{\alpha}$, which is indeed well realized by the data, shown in 
Fig. 1. 
The data plotted in the figure come from lattices with aspect ratio 
$L_1/L_2$ between $1$ and $3$ (see Ref.\cite{su2} for details).
Our assumption Eq. (\ref{plaqope}) is thus verified. The same analysis
was performed in Ref.\cite{z2} for $Z_2$ gauge theory.
\begin{center}
\begin{figure}[htb]
\vspace{4pt}
\mbox{\epsfig{file=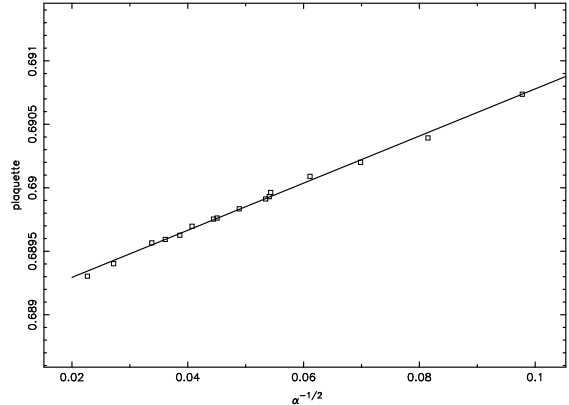}}
\caption{Time--like plaquette expectation value as a function of 
$\alpha^{-1/2}$ for $N_t=2$}
\end{figure}
\end{center}
\vskip0.3cm\noindent
Suppose we want to study the spatial distribution of the color flux tube at 
the deconfinement point in a static meson, that is in presence of two static
color sources. The correlator to be studied in $SU(2)$ theory is
\begin{eqnarray}
G(x,x_{1},x_{2})&=&\langle\Box(x) P(x_{1})P(x_{2}) \rangle\nonumber\\
&-&\langle \Box\rangle\langle P(x_{1})P(x_{2}) \rangle
\end{eqnarray}
where $P$ is the Polyakov loop operator. Universality tells us that at the
critical point
\begin{equation}
G(x,x_{1},x_{2})\propto \langle \epsilon(x) 
\sigma(x_{1})\sigma(x_{2})\rangle_{Ising}
\end{equation}
The Ising correlator on the r.h.s. is easily computed in conformal field 
theory and gives
\begin{equation}
G(x,x_{1},x_{2})\propto\frac{\left|x_{1}-x_{2}\right|^{3/8}}
{\left(\left|x-x_{1}\right|\left|x-x_{2}\right|\right)^{1/2}}
\end{equation}
which is represented in Fig.2.
\begin{figure}[htb]
\vspace{4pt}
\mbox{\epsfig{file=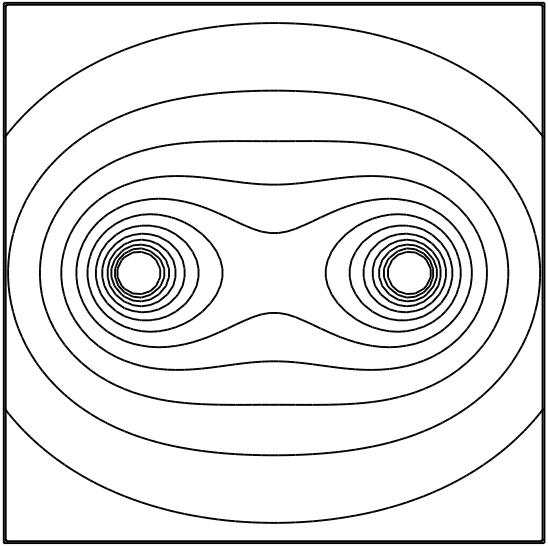}}
\caption{Color flux tube in a static meson}
\end{figure}
\vskip0.3cm\noindent
Also $(2+1)$-dimensional $SU(3)$ gauge theory can be studied within the
same approach. The Svetitsky-Yaffe conjecture predicts the universality
class to be the one of the two-dimensional three state Potts model,  
whose critical point is described by a minimal conformal field 
theory with central charge $c=4/5$.\\
In $SU(3)$ we can construct a static baryon by considering three static 
color sources. The color flux tube distribution is then given by the 
correlator:
\begin{eqnarray}
&&G(x,x_{1},x_{2},x_{3})=\langle\Box(x) P(x_{1})P(x_{2})P(x_{3}) \rangle
\nonumber\\
&&\ -\langle\Box \rangle\langle P(x_{1})P(x_{2})P(x_{3}) \rangle
\end{eqnarray}
which at the deconfinement point translates into a four point correlator
of the two-dimensional conformal theory:
\begin{equation}
G(x,x_{1},x_{2},x_{3})\propto \langle \epsilon(x) \sigma(x_{1})\sigma(x_{2})
\sigma(x_{3})\rangle_{c=4/5}
\end{equation}
This can be computed exactly and is expressed in terms of hypergeometric 
functions: We refer the reader to Ref.\cite{z2} for the analytical expression
and here we just show a plot of the flux tube distribution when the three
static sources are at the vertices of an equilateral triangle, Fig. 3.

\begin{figure}[htb]
\vspace{4pt}
\mbox{\epsfig{file=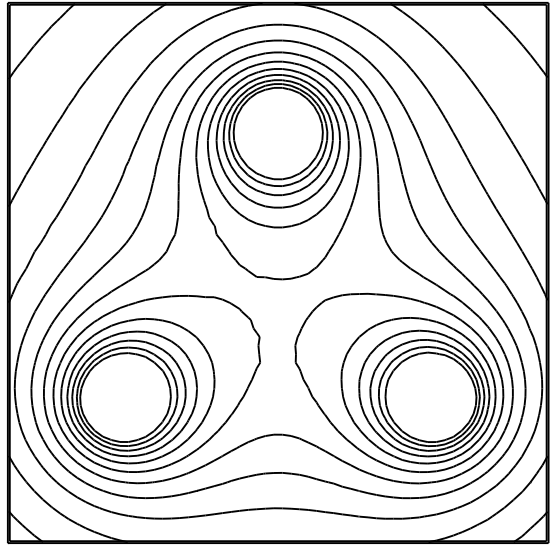}}
\caption{Color flux tube in a static baryon}
\end{figure}
\vskip0.3cm\noindent
In conclusion, two main lessons can be learned from our results:
\begin{itemize}
\item{Universality arguments provide a powerful, analytical approach to the 
non-perturbative physics 
of many interesting gauge theories.}
\item{This is especially true for $(2+1)$-dimensional gauge theories, since 
critical behavior 
in two dimensions is completely understood  with the methods of 
conformal field theory.}
\end{itemize}

\end{document}